\documentclass[twocolumn]{autart} 

\newcommand{\nocontentsline}[3]{}
\newcommand{\tocless}[2]{\bgroup\let\addcontentsline=\nocontentsline#1{#2}\egroup}

\usepackage[utf8]{inputenc}					
\usepackage[T1]{fontenc}					
\usepackage{ragged2e,anyfontsize}			
\usepackage{gensymb}						

\usepackage{makecell}
\usepackage{stackengine}

\usepackage{nicefrac}

\usepackage[acronym, nonumberlist, nogroupskip, nopostdot, nomain]{glossaries}
\usepackage{glossary-superragged}
\setglossarystyle{superragged}

\makeatletter
\patchcmd{\thenomenclature}
  {\@ifundefined{chapter}}
  {\@ifundefined{relax}}
  {}
  {}
\makeatother

\usepackage{commath} 

\usepackage{nccmath} 
\usepackage{color,soul} 
\usepackage{marvosym} 

\usepackage{bm} 
\usepackage{bbm} 
\usepackage{amssymb}%
\usepackage{pifont}
\usepackage{tensor}

\usepackage{array}

\usepackage[normalem]{ulem} 

\usepackage[flushleft]{threeparttable} 

\usepackage[position=bottom]{subfig}
\usepackage{graphicx} 						
\usepackage{multirow}                		
\usepackage{colortbl} 						
\usepackage[dvipsnames, table]{xcolor}				
\usepackage{flafter}						
\usepackage{float}			

\usepackage[section]{placeins} 

\usepackage{booktabs}

\usepackage{tcolorbox}
\tcbuselibrary{breakable}

\usepackage{bibentry} 

\usepackage{lipsum}

\usepackage[labelfont={bf},font=small]{caption} 

\usepackage{amsmath,amssymb,stmaryrd} 		
\usepackage{mathtools}						
\usepackage{textcomp}                 		
\usepackage{siunitx}						
\usepackage[version=3]{mhchem} 				

\usepackage{listings}

\definecolor{dkgreen}{rgb}{0,0.6,0}
\definecolor{gray}{rgb}{0.5,0.5,0.5}
\definecolor{mauve}{rgb}{0.58,0,0.82}

\usepackage{lipsum}							
\usepackage[shortlabels]{enumitem}			
\usepackage{pdfpages}						
\usepackage[footnote,draft,danish,silent,nomargin]{fixme}	
\usepackage{booktabs}

\usepackage{svg}
\usepackage{pgfplots} 
\pgfplotsset{compat=1.6}
\usepgfplotslibrary{groupplots}
\usetikzlibrary{decorations.markings,shapes.arrows}

\pgfplotsset{every axis/.append style={cycle list/Dark2}, ylabel near ticks, xlabel near ticks}
\pgfplotsset{every axis plot/.append style={very thick}}
\pgfplotsset{every axis legend/.append style={legend cell align=left,align=left, font=\small, fill opacity=0.7,draw opacity=1,text opacity=1}}
\pgfplotsset{every axis title/.append style={font=\itshape}}
\pgfplotsset{every major grid/.append style={dashed}}
\pgfkeys{/pgf/number format/.cd,1000 sep={}} 

\usepackage{tikz}
\usepackage{circuitikz}
\usepackage{pdfpages}
\usepackage{gensymb}
\usepackage{tabularx}
\usepackage{booktabs}
\usepackage{rotating} 
\usetikzlibrary{calc,patterns,angles,quotes,backgrounds,shapes, positioning,arrows,decorations,decorations.markings,shadows}
\usetikzlibrary{spy}

\ctikzset{tripoles/mos style/arrows}
\usepackage{tikz-timing}

\tikzset{
  declare function={
    atan3(\a,\b)=ifthenelse(atan2(0,1)==90, atan2(\a,\b), atan2(\b,\a));},
  kinky cross radius/.initial=+.125cm,
  @kinky cross/.initial=+, kinky crosses/.is choice,
  kinky crosses/left/.style={@kinky cross=-},kinky crosses/right/.style={@kinky cross=+},
  kinky cross/.style args={(#1)--(#2)}{
    to path={
      let \p{@kc@}=($(\tikztotarget)-(\tikztostart)$),
          \n{@kc@}={atan3(\p{@kc@})+180} in
      -- ($(intersection of \tikztostart--{\tikztotarget} and #1--#2)!%
             \pgfkeysvalueof{/tikz/kinky cross radius}!(\tikztostart)$)
      arc [ radius     =\pgfkeysvalueof{/tikz/kinky cross radius},
            start angle=\n{@kc@},
            delta angle=\pgfkeysvalueof{/tikz/@kinky cross}180 ]
      -- (\tikztotarget)}}}

\tikzset{
    partial ellipse/.style args={#1:#2:#3}{
        insert path={+ (#1:#3) arc (#1:#2:#3)}
    }
}

\definecolor{commentGreen}{RGB}{34,139,24}
\definecolor{stringPurple}{RGB}{208,76,239}

\definecolor{numbercolor}{gray}{0.7}		
\newif\ifchapternonum

\definecolor{tablecolor}{rgb}{0.89, 0.89, 0.89}
\newcommand{\units}[1]{[\si{#1}]}
\usepackage{caption,booktabs}

\usepackage{adjustbox}

\newcolumntype{?}{!{\vrule width 1pt}}
\allowdisplaybreaks
\begin{document}

\begin{frontmatter}

\title{Glucose-Insulin Dynamical Model for Type 2 Diabetic Patients} 

\author[Aalborg]{Mohamad Al Ahdab}\ead{Maah@es.aau.dk},    
\author[Aalborg]{John Leth}\ead{jjl@es.aau.dk},               
\author[Aalborg]{Torben Knudsen}\ead{tk@es.aau.dk},  
\author[Aalborg]{Henrik Clausen}\ead{hgcl@es.aau.dk} 

\address[Aalborg]{Aalborg University, Denmark}  

\begin{keyword}                           
T2D, Diabetes, Glucose, Insulin, Model               
\end{keyword}                             

\begin{abstract}                          
In this paper, a literature review is made for the current models of glucose-insulin dynamics of type 2 diabetes patients. Afterwards, a model is proposed by combining and modifying some of the available models in literature to take into account the effect of multiple glucose meals, multiple metformin doses, insulin injections, physical exercise, and stress on the glucose-insulin dynamics of T2D patients. The model is proposed as a candidate to be validated with real patients data in the future.
\end{abstract}

\end{frontmatter}

\section{Introduction}
One of the greatest health challenges which faces humanity in the 21st century is the emergence of type 2 diabetes (T2D) as a global pandemic. More than 415 million were reported to suffer from T2D in 2015 and the number is expected to reach 642 million by 2040 \cite{MoneyUSD}. Moreover, the global expenses related to T2D are estimated to be 850 million USD in 2017 and they are expected to increase \cite{pandemic}. T2D is characterized by high levels of glucose concentration in the blood. This increase in glucose levels can cause cardiovascular diseases and, if left untreated, will lead to organ failures. For T2D patients, low sensitivity to insulin, which is the hormone responsible for lowering glucose concentration in the blood, causes the beta cells in the pancreas to produce more of it to compensate. This will eventually weaken the cells and damage them which in turns will make the body fail to regulate glucose concentration \cite{AMCC}. Insulin based treatment is initiated at later stages of the T2D disease when changes in diets and physical activities accompanied with oral medications have failed. Clinically, it is difficult to calculate suitable insulin doses for each specific patient. Therefore, many patients experience uncontrolled hyperglycemia for a long period of time until they reach a safe level of glucose \cite{WONG2012169}. This happens due to the high variability of each patient and the different effect of their diet and lifestyle on insulin sensitivity. Furthermore, many patients suffer from hypoglycemic episodes if they overdose on insulin due to inaccurate calculations. Developing a model for the glucose-insulin dynamics helps with developing and testing algorithmic insulin dose guiders that handle patients variability better and ensure a safe reach for a desired blood glucose concentration. Moreover, having a model can helps in the development of hypoglycemic episodes alert systems for the patients. Additionally, models for glucose-insulin dynamics can help medical professionals with their patients' treatment plans.

In this paper, a survey of current models and simulators for the insulin-glucose dynamics in T2D patients is carried out. After that, a new model is proposed by combining and modifying some of the models from current literature. Finally, simulation results are provided to discuss the introduced model.

\section{Literature Review}
In general, there are two main categories of methods to model systems: first principles methods, or data driven methods derived by fitting data to general mathematical structures such as ARMAX models. 

The glucose-insulin dynamical models for T2D patients based on first principles can vary with different degrees of complexity. Generally in the literature, there exist two main categories of such models: minimal models and maximal models \cite{cobelli2009diabetes}.
Maximal models are very detailed models which model metabolic functions at a molecular level. On the other hand, minimal models are less detailed and rely mostly on compartments and mass balance equations. While maximal models provide a great level of accuracy, the amount of different data which is required to estimate parameters for the models is large and difficult to obtain from patients undergoing typical treatment plans.
Moreover, the high accuracy of maximal models provides little relevance to the accuracy of the general glucose-insulin dynamics within the human body \cite{cobelli2009diabetes}.

In contrast, minimal models consist of compartments to represent the distribution, diffusion, and production of glucose and insulin in the body with terms to represent the interaction between them. Furthermore, these models include pharmacokinetic equations to describe exogenous insulin injections and the intake of other medications. Several simulation oriented models for type 1 diabetic (T1D) patients were developed such as the ones in \cite{Dalla}\cite{knad}\cite{Hovorka_2004}. The models for T1D patients can be extended to model T2D patients by including dynamics for endogenous insulin production and by obtaining new probability distributions for their parameters with the use of T2D patients data. 
As for already existing T2D models, Cobelli's model in \cite{Cobelli} is a detailed model for glucose-insulin dynamics which is intended for simulation use. It includes nonlinear terms with hybrid differential equations. The model parameters are mainly estimated from people who do not suffer from diabetes. However, general parameters for T2D patients were also estimated using data from T2D patients. Although the model does not consider molecular level dynamics, its level of complexity still requires a variety of measured variables to estimate parameters for a given patient. In addition to glucose measurement in the plasma, the model used traced consumed glucose data, plasma insulin concentration measurements, and data regarding C-peptides (Amino Acids molecules which are side product of insulin secretion). The authors in \cite{Cobelli} provided only mean parameters for T2D patients. They later developed an official simulator based on an extended version of their model to account for oral medications, Glucagon, and physical activities. The simulator is reported to have joint probability distribution for the parameters of T2D patients \cite{PadovaUVA}. Nevertheless, these distributions were not published. Moreover, there are no published material for the equations and the structure of the extended model.
Another detailed model is recently developed by \cite{Vahili}. It is an extension from the model in \cite{sorensen1985physiologic}. The model includes the affect of oral medications, Glucagon, and Glucagon-like peptide-1. The model is provided with mean parameters. Some of the parameters are identified from clinical data, while others are taken from different sources of literature. None of the mentioned T2D models so far considers or provide a mathematical structure for injected insulin, ingestion of multiple meals, and physical activity.
 A simpler T2D simulation oriented model with insulin injection is provided in \cite{Rikke}. It was developed as an extension to the model in \cite{Jaus} to take into account when both fast acting and long acting insulin are used for treatment. The model also uses modulator functions to account for the circadian rhythm in the glucose-insulin dynamics. As for parameter estimation, the model requires data of consumed glucose, glucose concentration in the plasma, injected insulin, and endogenous insulin concentration to be identifiable. In \cite{Rikke}, parameter estimation was carried out using data from two different trials with different types of insulin with a total of 29 T2D patients. The parameter estimation took into account inter-individual variety by estimating a mean and a variance for the parameters. Therefore, the model can be used to simulate a population of T2D patients. The model, however, is simple and does not include oral medications, physical activity, and glucose ingestion.
Another model is the one from \cite{ARADOTTIR201715086}. This model requires plasma glucose measurement with injected insulin concentration data. Therefore, this model can be validated, improved, and fitted with data collected from patients during typical treatment plans as done by the work in []. The model takes the T1D model in \cite{knad} and extended it with a term to describe endogenous insulin production. The model, nevertheless, does not include a mathematical structure for physical activity or oral medications. Probability distribution for the model parameters can be found in \cite{ARADOTTIR201715086}.

For data driven models, several methods were attempted for T1D in  \cite{T1DTimes1,ChaoticT1D,valletta} and T2D in \cite{Barbra}. These models were generally developed to predict and detect hypoglycemic risks. However, data driven models are difficult for the purpose of developing a general model for T2D patients simulations. One major problem is that when data is collected from patients, the patients are already following some form of a feedback control mechanism between measured glucose levels and injected insulin. Therefore, fitting models on data from insulin inputs to glucose output will also include the dynamics of the feedback mechanism. This is undesired when the model is intended to test different control algorithms or if it is intended for the development of control strategies. This problem can be solved by perturbing the input insulin doses with time varying functions. However, it is difficult to impose perturbation on the insulin input for the patients without affecting their comfort and health.

In this work, it is intended to provide a model with mathematical structures for the effect of multiple glucose meals, insulin injections, multiple oral doses of metformin with different sizes, physical activity, and stress. The model is based on the one in \cite{Vahili} with modifications and inclusions as following (see figure \ref{fig:sum}):
\begin{itemize}
    \item Modifying the model to account for multiple meals (see section \ref{sec:Glucose_Absorption_Model}).
    \item Including a model for insulin injections based on \cite{li2009mathematical} (see section \ref{sec:InsInj_subsystem}).
    \item Modifying the metformin model to account for multiple different doses in section \ref{sec:Metformin}.
    \item Including the effect of physical activity based on \cite{Breton2008} (see section \ref{sec:Physical_Activity}).
    \item Including the effect of stress based on \cite{Stress} (see section \ref{sec:Sress}).
\end{itemize}
 In addition, two simulation cases are provided in section \ref{sec:SimR} to demonstrate the effect of lifestyle changes with the model.
 The model proposed in this paper is intended to be used as a starting point for developing an open source simulator for T2D patients when data is available. 
 All the simulations in the paper are done with a Matlab code which can be obtained upon request from the authors. The model parameters which are used in the simulations are found in \ref{tab:Param}.
 
\section{Model Description}
\begin{figure*}[h]
    \centering{\resizebox{0.95\textwidth}{!}{\input{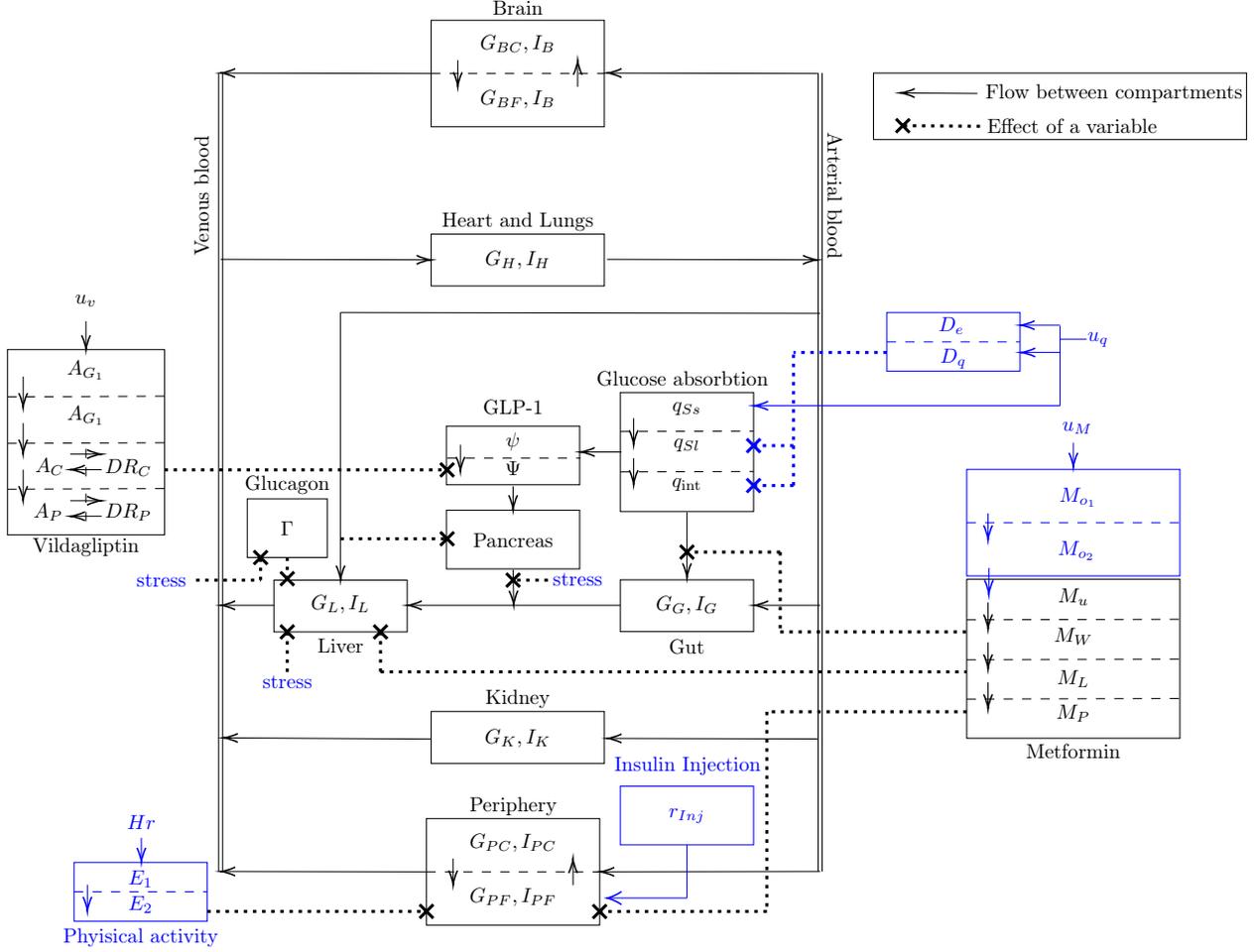}}}    \caption{A summary of the overall model with blue indicating the modified or added models compared to \cite{Vahili}}
    \label{fig:sum}
\end{figure*}

The model is mainly based on the one from  \cite{Vahili} with the following four main subsystems:\begin{itemize}
    \item Glucose subsystem.
    \item Insulin subsystem. 
    \item Glucagon subsystem.
    \item Incretins hormone subsystem.
\end{itemize}
See figure \ref{fig:sum} for an overview of the model. The glucose and insulin subsystems are modelled as a set of compartments representing different main parts of the human body: brain, heart and lungs, guts, liver, kidney, and peripherals. The flow between these compartments follows the human blood cycle. As for the glucagon and the incretins, a single compartment is used for each one of them as it is assumed that glucagon and incretins have equal concentration in all the body parts. In addition, the model contains metabolic production and uptake rates for different compartments. These metabolic rates are generally defined as their basal values multiplied with scaling variables that depend on the concentrations of insulin, glucose, and/or glucagon (see \eqref{eq:rdefine}).
The pancreas has a different nonlinear and hybrid model. In addition, a glucose ingestion model based on \cite{Cobelli} is included as in \cite{Vahidi2016} but modified to handle multiple meals along the day. Moreover, metformin and vildagliptin oral treatment models are included based on \cite{Sun2011} and \cite{Landersdorfer2012} respectively as in \cite{Vahili} but with a modification on the oral metformin model to handle different oral doses along the treatment. Additionally, a physical activity model based on \cite{Breton2008} is added to the model. Furthermore, long acting and fast acting insulin injection models based on \cite{li2009mathematical} are added. Finally, stress is included as a factor $\alpha_{s}\in[0,1]$ as in \cite{Stress}. The main model includes parameters that were estimated by \cite{sorensen1985physiologic} for a healthy 70 $\si{kg}$ male. The work in \cite{Vahidi2016} considered a subset of these parameters to be estimated for the diabetic cases. Parameters for the different added models are taken from their corresponding literature.
In the following subsections, the added and modified models and states will be discussed. The full model equations are provided in appendix \ref{sec:FullModelEq}.

\subsection{Glucose Absorption Model}
\label{sec:Glucose_Absorption_Model}
In this section, a modification is introduced to account for multiple glucose meal sizes. 
The model used for glucose absorption in \cite{Vahili} considers only one glucose meal and was used for oral glucose tests where the patient were given an oral glucose dose and asked to fast while data is collected. The model is given as:
\begin{subequations}
\label{eq:Glucose-Absorption}
\begin{align}
    \label{eq:qSs}
    &\frac{dq_{Ss}}{dt}=-k_{12q}q_{Ss}\\
    &\frac{dq_{Sl}}{dt}=-k_{\mathrm{empt}}q_{Sl}+k_{12q}q_{Ss}\\
    &\frac{dq_{\mathrm{int}}}{dt}=-k_{\mathrm{abs}}q_{\text{int}}+k_{\mathrm{empt}}q_{Sl}\\
    \label{eq:kempt}
    \begin{split}
    &k_{\mathrm{empt}}=
        k_{\min }+\frac{k_{\max }-k_{\min }}{2}\left\{\vphantom{\int_0.8^2}\right.\\
        &\qquad\left.\tanh \left[\varphi_{1}\left(q_{S s}+q_{S l}-k_{\varphi_{1}} D_{q}\right)\right]\right.\\
        &\qquad\left.-\tanh \left[\varphi_{2}\left(q_{S s}+q_{S l}-k_{\varphi{2}} D_{q}\right)\right]+2\right\}\end{split}\\
    &\varphi_{1}=\frac{5}{2D_{{q}}(1-k_{\varphi_{1}})}\\
    &\varphi_{2}=\frac{5}{2D_{{q}}(k_{\varphi_{2}})}\\
    &Ra=f_{q}k_{\mathrm{abs}}q_{\mathrm{int}}
\end{align}
\end{subequations}
Where $q_{Ss}(0)=D_{q}~\units{mg}$ is the oral glucose quantity, $Ra$ is the rate of glucose appearance in the blood, $f_{q}$ is an absorption factor, $k_{12}~\units{min^{-1}}$ and $k_{\mathrm{abs}}~\units{min^{-1}}$ are the rate constants for glucose transfer to stomach and glucose absorption in the intestines respectively, $k_{\mathrm{empt}}~\units{min^{-1}}$ is a rate parameter for emptying the stomach of glucose to the intestines. This parameter can have values between $k_{\mathrm{min}}$ and $k_{\mathrm{max}}$ depending on the glucose dose size $D_{q}$. In order to make the model handle different meals with different time instants, the parameter $D_{q}$ needs to be modified according to the meal sizes and time. The following are the proposed modifications:
\begin{subequations}
\label{eq:Glucose-Absorption}
\begin{align}
    \label{eq:qSs}
    &\frac{dq_{Ss}}{dt}=-k_{12q}q_{Ss}+\sum_{i=1}^{N_{q}(t)}u_{q_{i}}\delta(t-t_{i}), ~ i\in\mathbb{Z}_{+}\\
    &\frac{d D_{e}}{dt}= -k_{q}D_{e}+\sum_{i=1}^{N_{q}(t)}u_{q_{i}}\delta(t-t_{i})\\
    &\frac{d D_{q}}{dt}= k_{q}\left(u_{q_{N_{q}(t)}}-D_{q}\right)+D_{m}\sum_{i=1}^{N_{q}(t)}\delta(t-t_{i})\\
    &D_{m}=\begin{cases}
    D_{e}-D_{q} & u_{q_{N_{q}(0)}} \neq 0\\
    1 & u_{q_{N_{q}(0)}}=0
    \end{cases}
\end{align}
\end{subequations}
Where $q_{Ss}(0)=0$, $D_{e}(0)=0$, $\delta(t-t_{i})$ is the Dirac delta distribution, $t_{i}$ is the time instance for meal $i$, $N_{q}(t)$ is the integer number of meals until time $t$, $u_{q_{i}}~\si{[mg]}$ is the amount of oral carbohydrates intake for meal $i$. The state $D_{e}$ is introduced to handle the accumulation of carbohydrates meals with a decay factor $k_{q}~\si{[min^{-1}]}$ in order to remove the effect of meals with time. With that, parameter $D_{q}$ is now a state updated by $D_{e}$ each time a new meal is consumed and made to converge to the last given meal amount $u_{q_{i}}$ with the same rate factor $k_{q}$ such that it converges to the original model through time if no meal is consumed afterwards. 
Note that $D_q(0)=0$ when a zero carbohydrates meal ($u_{q_{N_{q}(0)}}=0$) is assumed at time $t=0$, which leads to \eqref{eq:kempt} being undefined ($\varphi_{1}q_{S s}=\infty \,0$). To avoid this, the state $D_{m}(0)$ is set to $1$ when $u_{q_{N_{q}(0)}}=0$.
Note that the value $D_{m}(0)$ can have any nonzero value in the case of zero carbohydrates meal at $t=0$. The value $D_{m}(0)$ will not affect the rate of glucose appearance in the plasma since the states $q_{Ss}$ for ingested carbohydrates, and $D_{e}$ for the effect of accumulation of meals depend on $u_{q_{N_{q}(0)}}$ and not $D_{m}$.
Parameters $f_{q},~k_{\varphi_{1}},$ and $k_{\varphi_{2}}$ are known and taken from \cite{Cobelli}. The rest of the parameters, $k_{12q},~k_{\min},~k_{\max},k_{\mathrm{abs}}$, and $k_{q}$ are to taken to be the mean parameters which were estimated in \cite{Vahidi2016}. The introduced parameter $k_{q}$ has no estimate. Therefore, it is assumed to be equal to $k_{\mathrm{min}}$.

\begin{figure}[h!]
    \centering
    \includegraphics[width=0.47\textwidth]{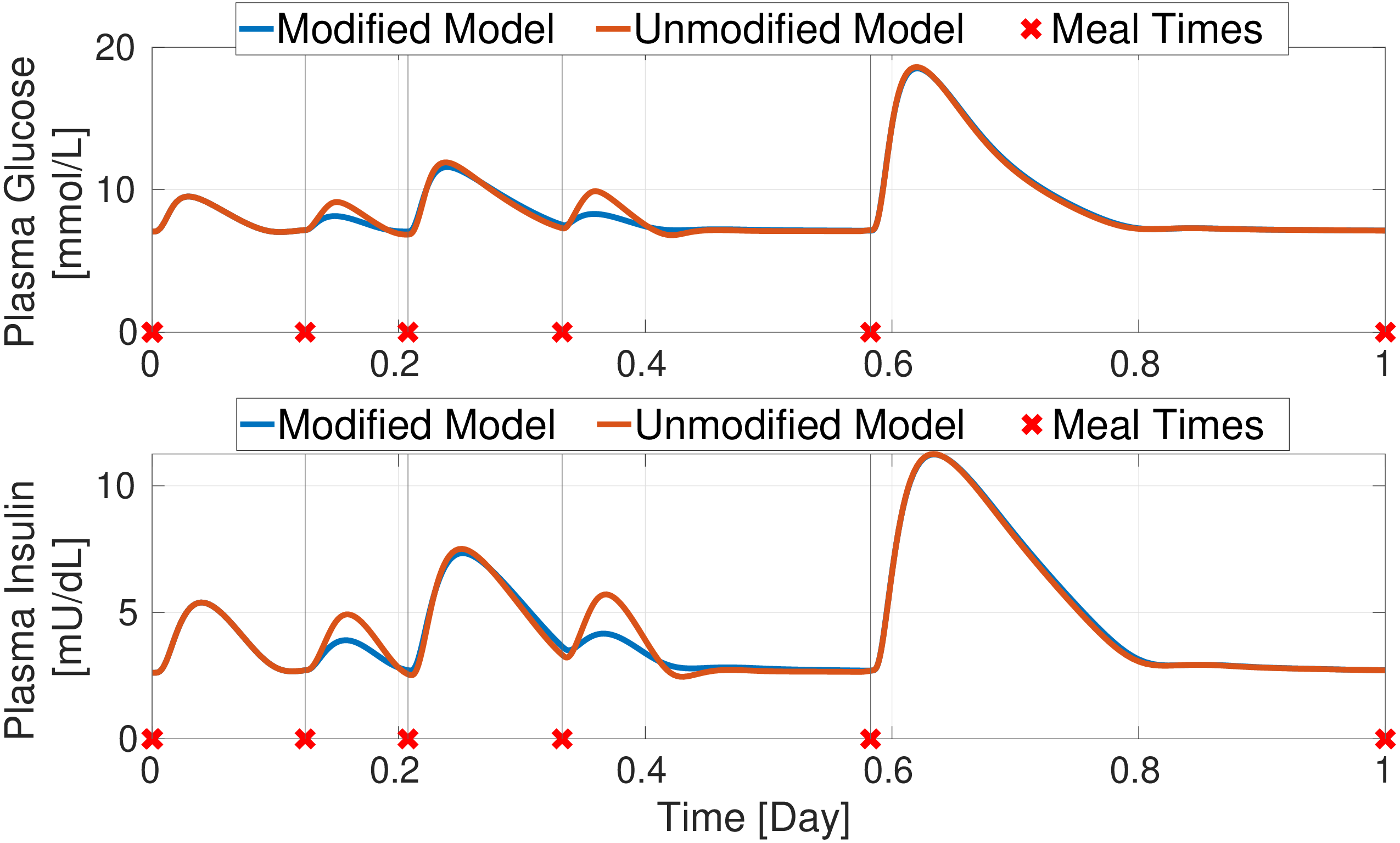}
    \caption{Simulation results for the modified glucose absorption model against the unmodified one.}
    \label{fig:Meals}
\end{figure}

A simulation of a patient with the modified meals model compared against the unmodified one is shown in Figure \ref{fig:Meals}. The patient is consuming a breakfast meal of $30~\units{g}$ carbohydrates, a second breakfast meal of $10~\units{g}$ carbohydrates, a lunch meal of $50~\units{g}$ carbohydrates, an afternoon snack of $10~\units{g}$ carbohydrates, and a dinner meal of $110~\units{g}$ carbohydrates. The simulated patient has a basal value of $G_{PC}(0)=8~\units{mmol\,L^{-1}}$ for glucose concentration in the central periphery compartment and $I_{PF}=1~\units{mU\,L^{-1}}$ for the insulin concentration in the interstitial fluid periphery compartment.
It can be seen from the simulation results that the glucose appearance in plasma is distributed in a larger window of time with lower peaks for meals that are close to each other. This is due to the reduction of the stomach emptying rate $k_{\mathrm{empt}}$ in response to increased accumulation of ingested carbohydrates captured by the state $D_{q}$. Additionally, glucose appearance in plasma for the modified model in response to meals after hours of fasting closely resembles the glucose appearance in plasma for the unmodified model as can be seen for the dinner and breakfast meal. This is intended since the unmodified model was proposed for fasting conditions.

\subsection{Insulin Injection Model}
\label{sec:InsInj_subsystem}
\begin{figure}[h!]
    \centering
    \includegraphics[width=0.45\textwidth]{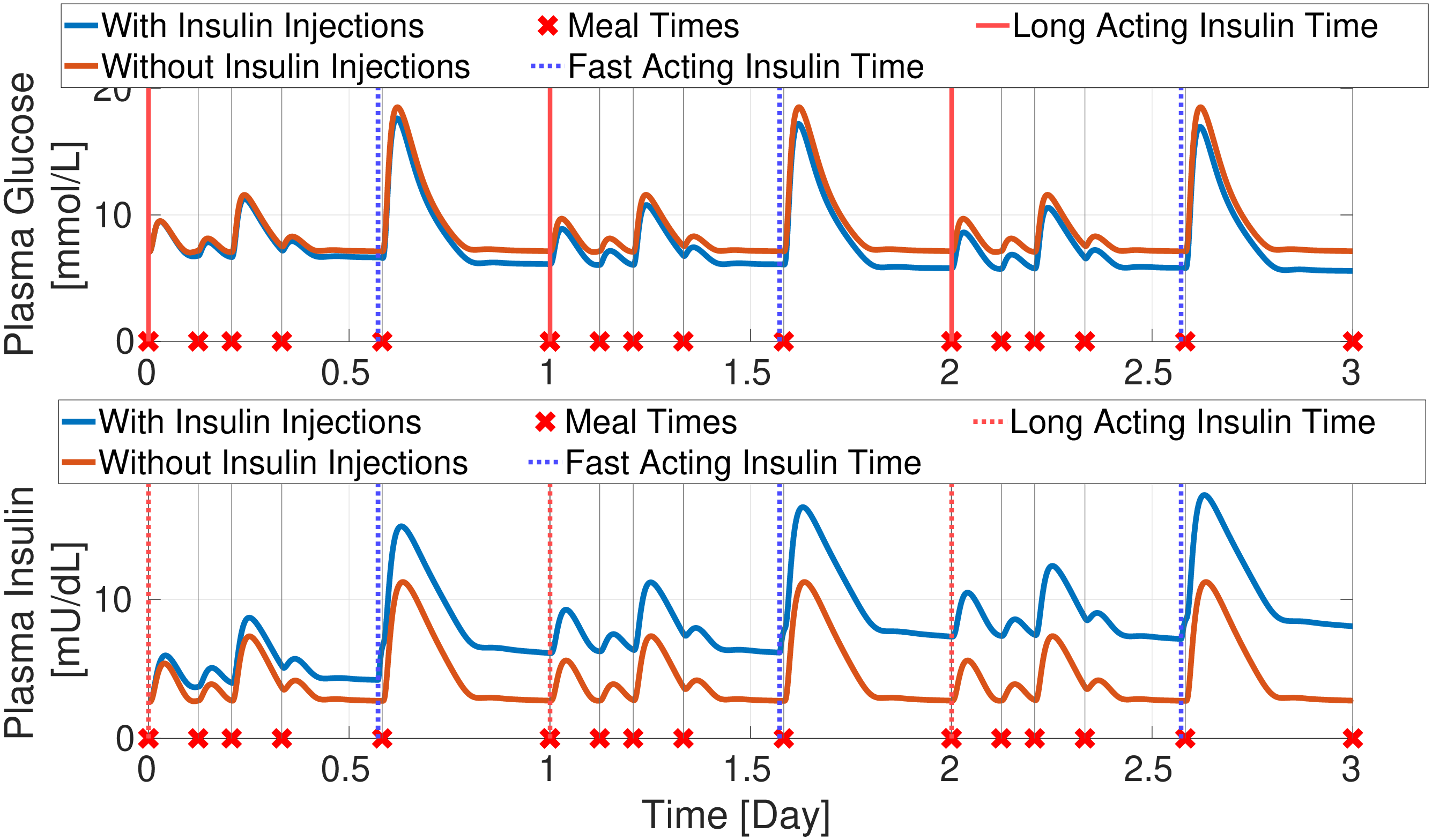}
    \caption{A simulation showing the effect of injected insulin on glucose and insulin concentrations.}
    \label{fig:InsulinFig}
\end{figure}

In this section, a model for long acting and fast acting insulin injections based on the one from \cite{li2009mathematical} is introduced in \cite{Vahili}.
Both fast and long acting insulin analogues treatments are considered for the model.
When analogue insulin is injected, it dissociates from its hexameric form to dimers and monomers which then can penetrate the capillary membrane and get absorbed into the plasma. For fast acting insulin, only two compartments are considered: a compartment for insulin in its hexameric form, and a compartment for insulin in its dimeric and monomeric form. The following are the equations for fast acting insulin:
\begin{subequations}
\label{eq:fast-acting}
\begin{align}\begin{split}
&\frac{dH_{fa}}{dt}=\sum_{i=1}^{N_{fa}(t)}\delta\left(t-t_{i}\right)\frac{10}{V^{I}_{PF}}u_{f_{i}}\\
&\qquad-p_{fa}\left(H_{fa}(t)-q_{fa}D_{fa}^{3}(t)\right)\end{split} \\
&\frac{dD_{fa}}{dt}=p_{fa}\left(H_{fa}(t)-q_{fa}D_{fa}^{3}(t)\right)-\frac{b_{fa} D_{fa}(t)}{1+I_{PF}(t)}
\end{align}
\end{subequations}
Where $H_{fa}~[\si{mU\,dL^{-1}}]$ is the concentration of injected fast acting insulin in its hecameric form, $D_{fa}~~[\si{mU\,dL^{-1}}]$ is the concentration of insulin in its diameric and monomeric form, $N_{fa}(t)$ is the number of injected fast acting insulin doses until time $t$, $u_{f_{i}}~\units{mU}$ is the amount of injected fast acting insulin, $b_{fa}~[\si{min^{-1}]}$ is a constant for the infusion of fast acting insulin into the body, $p_{fa}~\units{min^{-1}}$ is a constant diffusion parameter, $q_{fa}~\units{dL^2\,mU^{-2}}$ is a constant such that $p_{fa}q_{fa}$ is the parameter for fast acting insulin dimers converting back to hexamers, and $I_{PF}~\units{mU\,dL^{-1}}$ is the insulin concentration in the interstitial periphery compartment.
Parameters for Lispo and Aspart insulin injection are reported in \cite{li2009mathematical}.
For long acting insulin, an extra state $B_{la}$ is added in \cite{li2009mathematical} to take into account the increased delay in the dissociation of hexameric insulin to dimers and monomers:
\begin{subequations}
\label{eq:Long-acting}
\begin{align}
&\frac{dB_{la}}{dt}=\sum_{i=1}^{N_{la}(t)}\delta\left(t-t_{i}\right)\frac{10}{V^{I}_{PF}}u_{l_{i}}-k_{la}B_{la} \frac{C_{\max }}{1+H_{la}}\\
&\frac{dH_{la}}{dt}=k_{la}B_{la} \frac{C_{\max }}{1+H_{la}}-p_{la}\left(H_{la}-q_{la}D_{la}^{3}\right) \\
&\frac{dD_{la}}{dt}=p_{la}\left(H_{la}-q_{la}D_{la}^{3}\right)-\frac{b_{la} D_{la}}{1+I_{PF}}
\end{align}
\end{subequations}
Where $B_{la}~\units{mU\,dL^{-1}}$ is the added bound state for the concentration of hexameric insulin before diffusing, $H_{la}~[\si{mU\,dL^{-1}}]$ is the concentration of injected long acting insulin in its hexameric form, $D_{la}~[\si{mU\,dL^{-1}}]$ is the concentration of injected insulin in its diameric and monomeric form, $N_{la}(t)$ is the number of injected long acting insulin doses until time $t$, $u_{l_{i}}~\units{mU}$ is the amount of long acting insulin dose at time $t_{i}$, $b_{la}~[\si{min^{-1}]}$ is a constant for the infusion of long acting insulin into the body, $p_{la}~\units{min^{-1}}$ is a constant diffusion parameter for long acting insulin, $q_{la}~\units{dL^2\,mU^{-2}}$ is a constant such that $p_{la}q_{la}$ is the parameter for long acting insulin dimers converting back to hexamers, $k_{la}~\units{min^{-1}}$ is a constant absorption rate, and $C_{\max}$ is a dimensionless maximum transmission capacity constant. Parameters for insulin Glargin are reported in \cite{li2009mathematical}.
The injected insulin enters the interstitial periphery compartment \eqref{eq:IPF} with the following rate $r_{inj}$: 
\begin{equation}
    \label{eq:Inject_Insulin_PH}
        r_{Inj}=V^{I}_{PF}\frac{r_{la}b_{la}D_{la}}{1+I_{PF}}+V^{I}_{PF}\frac{r_{fa}b_{fa}D_{fa}}{1+I_{PF}}
\end{equation}
Where $r_{la},r_{fa}\leq1$ are the fractions of long acting and fast acting insulin that get to the periphery compartment, and $V^{I}_{PH}~\units{L}$ is the volume of the interstitial compartment. Figure \ref{fig:InsulinFig} shows a simulation for a patient having the same basal values and following the same meal plan as the simulation discussed in section \ref{sec:Glucose_Absorption_Model}. The patient takes a long acting insulin dose of $50~\units{U}$ everyday an hour before the breakfast meal. Additionally, the patient takes a $30~\units{U}$ of fast acting insulin 15 minutes before dinner. Long acting insulin lower the glucose concentration over a large window of time. Moreover, the fast acting insulin helps at reducing the glucose peak after dinner.

\subsection{Metformin}
\label{sec:Metformin}
\begin{figure}[h!]
    \centering
    \includegraphics[width=0.45\textwidth]{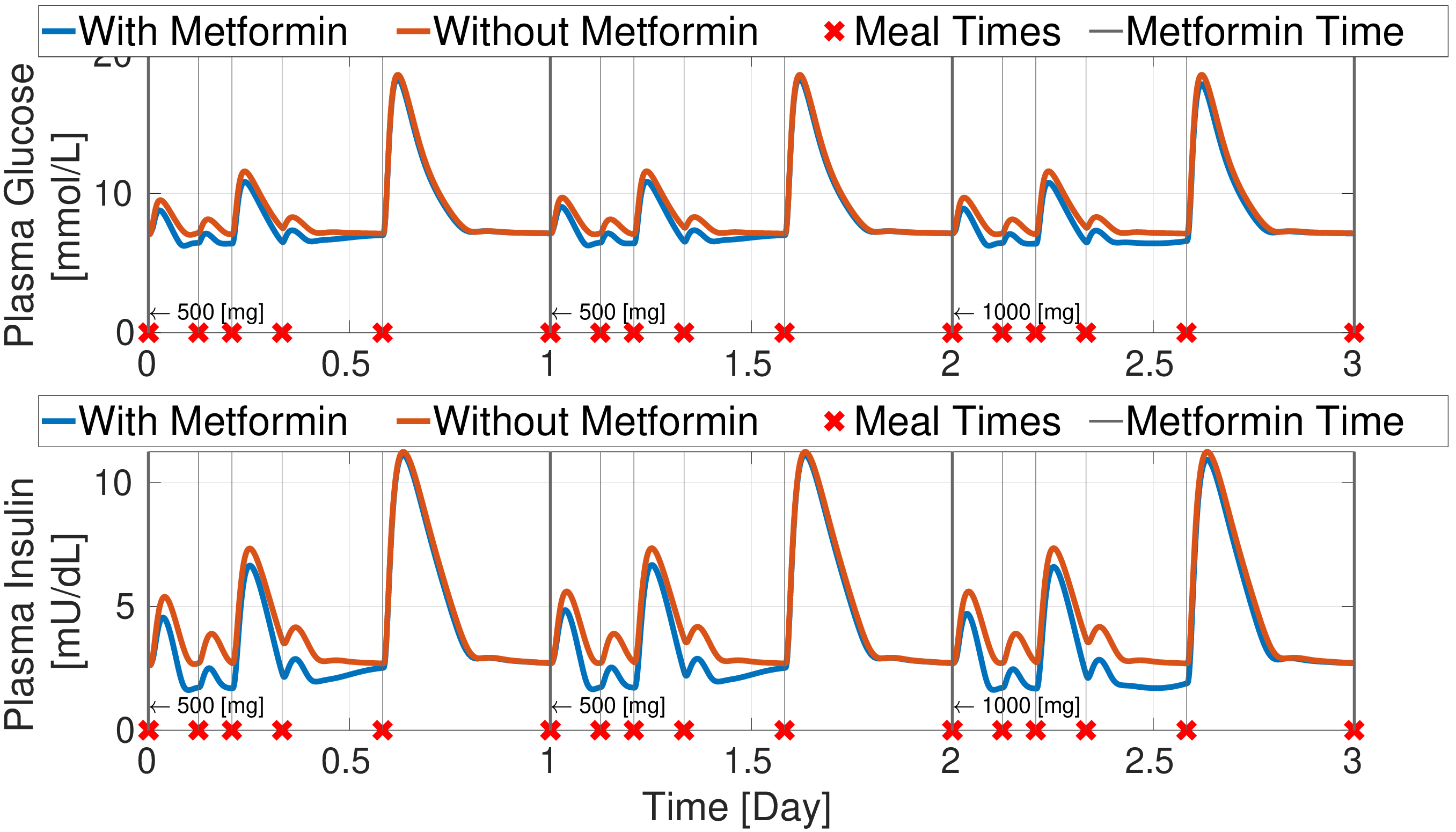}
    \caption{A simulation showing the effect of metformin on glucose and insulin concentrations.}
    \label{fig:Metforminplot}
\end{figure}

In this section, a modification for the metformin model in \cite{Vahili} is carried out to account for multiple doses of oral metfromin with different amounts.
The metformin model used in \cite{Vahili}, including the pharmacokinetic and its interaction with full glucose-insulin dynamical models, is based on \cite{Sun2011}.
The pharmacokinetic model of metformin in \cite{Sun2011} is given as following:
\begin{subequations}
\label{eq:PK-Metaformin}
\begin{align}
\frac{dM_{GL}}{dt}&=-M_{GL}\left(k_{g o}+k_{g g}\right)+M_{O}\\
\frac{dM_{GW}}{dt}&=M_{GL} k_{g g}+M_{P} k_{p g}-M_{GW} k_{g l}\\
\frac{dM_{L}}{dt}&=M_{GW} k_{g l}+M_{P} k_{p l}-M_{L} k_{l p}\\
\frac{dM_{P}}{dt}&=M_{L} k_{l p}-M_{P}\left(k_{p l}+k_{p g}+k_{p o}\right)+M_{GL}
\end{align}
\end{subequations}
Where $M_{GL}~\units{\mu g}$ is the metformin amount in the gastrointestina lumen, parameters $k_{g o},k_{g g},k_{p g},k_{g l},k_{p l}$,
$k_{l p},k_{p o}~\units{min^{-1}}$ are transfer rate constants between the compartments, and $M_{O}$ is the flow rate of orally ingested metformin which is modelled as:
\begin{equation}
    \label{eq:MO}
    M_{O}=Ae^{-\alpha_{M}t}+Be^{-\beta_{M}t}
\end{equation}
Where $A,B~\units{\mu g\,min^{-1}}$ and $\alpha_{M},\beta_{M}~\units{min^{-1}}$ are constant parameters that were identified in \cite{Sun2011}. These parameters were identified with data in which patients were taking only a $500~\units{mg}$ oral dose of metformin. Therefore, the model is modified in this work to take into account different amount of doses at different times by introducing the following:
\begin{subequations}
\begin{align}
    \frac{dM_{O1}}{dt}&=-\alpha_{M}M_{O1}+\sum_{i=1}^{N_{M}(t)}\delta\left(t-t_{i}\right)u_{M_{i}}\\
    \frac{dM_{O2}}{dt}&=-\beta_{M}M_{O2}+\sum_{i=1}^{N_{M}(t)}\delta\left(t-t_{i}\right)u_{M_{i}}\\
    M_{O}&=\rho_{\alpha}M_{01}+\rho_{\beta}M_{02}
\end{align}
\end{subequations}
With $N_{M}(t)$ being the number of consumed doses of metformin until time $t$, $u_{M_{i}}~\units{\mu g}$ is the amount of metformin consumed at time $t_{i}$, and the constants  $\rho_{\alpha}=A/(500000~\units{\mu g})~\units{min^{-1}}$ and $\rho_{\beta}=B/(500000~\units{\mu g})~\units{min^{-1}}$ are rate parameters. Figure \ref{fig:Metforminplot} shows a simulation result of the same patient in section \ref{sec:Glucose_Absorption_Model} taking metformin doses of $500~\units{mg}$ for the first two days and then a metformin dose of $1000~\units{mg}$ for the last two days. The $1000~\units{mg}$ dose prolong the effect of metformin on lowering the glucose concentration when compared to the dose of $500~\units{mg}$.
\subsection{Physical Activity Model}
\label{sec:Physical_Activity}
\begin{figure}
    \centering
    \includegraphics[width=0.45\textwidth]{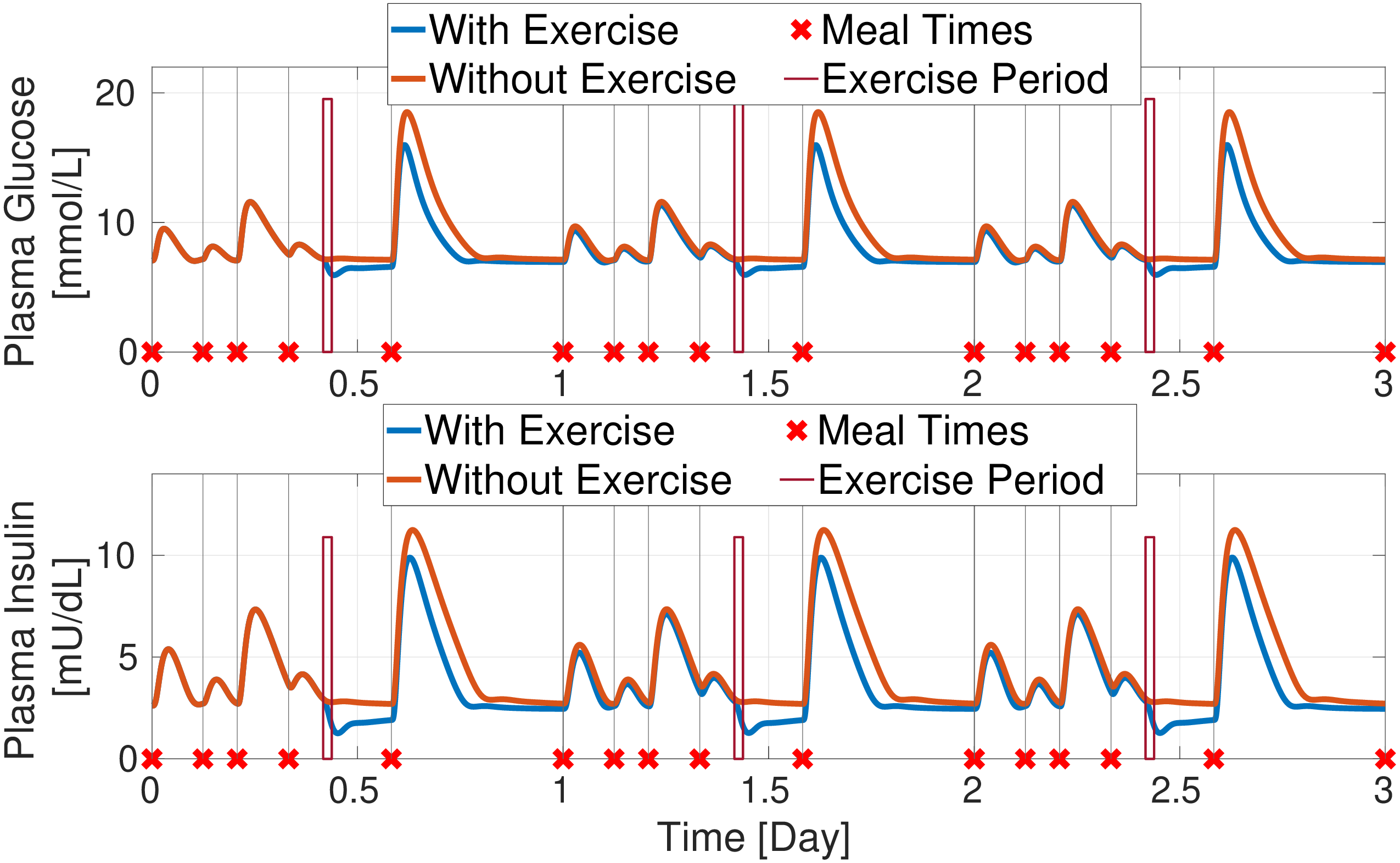}
    \caption{A simulation showing the effect of physical activity on glucose and insulin concentrations.}
    \label{fig:pha}
\end{figure}

In this section, a physical activity model based on \cite{Breton2008} is added to the model in \cite{Vahili}.
The model in \cite{Breton2008} was developed for a T1D model based on \cite{Bergman1979}. The model considers the change of the heart beat following a physical activity to be the stimulus of two states $E_{1}$ and $E_{2}$ which are dimensionless:
\begin{subequations}
\begin{align}
\label{eq:Physical_activity_model}
\frac{dE_{1}}{dt}&=-\frac{1}{\tau_{\mathrm{HR}}}E_{1}+\frac{1}{\tau_{\mathrm{HR}}}\left(\mathrm{HR-HR}_{b}\right)\\
\frac{dE_{2}}{dt}&=-\left(g_{e}(E_{1})+\frac{1}{\tau_{e}}\right)E_{2}+g_{e}(E_{1})\\
g(E_{1})&=\frac{\left(\frac{E_{1}}{a_{e}\mathrm{HR}_{b}}\right)^{n_{e}}}{1+\left(\frac{E_{1}}{a_{e}\mathrm{HR}_{b}}\right)^{n_{e}}}
\end{align}
\end{subequations}
Where $t_{\mathrm{HR}},~\tau_{e}~\units{min}$ are time constants, $\mathrm{HR},~\mathrm{HR}_{b}~\units{bpm}$ are the current and rest heart rates respectively, and the parameters $a_{e},~n_{e}$ are dimensionless parameters. The first state $E_{1}$ is used directly as a stimulus to increase the insulin-independent glucose uptake in response to a physical activity while the state $E_{2}$ is used for the longer lasting change of insulin action on glucose. The glucose and insulin model structure in \cite{Breton2008} is simpler than the one considered in this work. Nevertheless, the inclusion of the physical activity for the model in this work is similar to how other models include physical activity, e.g., see \cite{DallaMan2009Phys}. With that, the effect of the state $E_{1}$ is included as an increase in the clearance rate of glucose in the periphery interstitial fluid compartment with a constant parameter $\beta_{e}~\units{bpm^{-1}}$ as  $\frac{1}{T^{G}_{P}}\left(1+\beta_{e}E_{1}\right)$ where $\frac{1}{T^{G}_{P}}~\units{min^{-1}}$ is the clearance rate for glucose in the periphery interstitial fluid compartment in \eqref{eq:GPF}. The effect of $E_{1}$ can be removed to obtain the original model by setting $\beta_{e}=0$.
As for the effect on insulin action, the state $E_{2}$ is introduced on the glucose metabolic rates which depend on insulin as following:
\begin{itemize}
    \item An increase in the periphery glucose uptake rate $r_{PGU}$ in the the interstitial fluid periphery compartment \eqref{eq:GPF} by a constant $\alpha_{e}$ as $\left(1+\alpha_{e}E_{2}\right)r_{PGU}$.
    \item An increase in the hepatic glucose uptake rate $r_{HGU}$ in the liver compartment \eqref{eq:GL} with a constant $\alpha_{e}$ as $\left(1+\alpha_{e}E_{2}\right)r_{HGU}$.
    \item A decrease in the hepatic glucose production rate $r_{PGH}$ in the liver compartment \eqref{eq:GL} with a constant $\alpha_{e}$ as $\left(1+\alpha_{e}E_{2}\right)r_{HGP}$.
\end{itemize}
The effect of $E_{2}$ can be removed to obtain the original model by setting $\alpha_{e}=0$.
The parameters for the physical activity model are taken from \cite{Breton2008} except for $\alpha_{e}$ and $\beta_{e}$ which were tuned to have a similar effect to the ones demonstrated in \cite{Breton2008,DallaMan2009Phys}. Figure \ref{fig:pha} shows a simulation for the patient described in section \ref{sec:Glucose_Absorption_Model} when the patient exercise everyday before dinner raising the heart rate from a rest value of $80~\units{bpm}$ to a value of $140~\units{bpm}$ for 30 minutes. The immediate effect of physical activity is seen in the simulation results. In addition, the prolonged effect of physical activity on insulin action on glucose is seen in the figure.  
\subsection{Stress Effect}
\label{sec:Sress}
\begin{figure}[h!]
    \centering
    \includegraphics[width=0.47\textwidth]{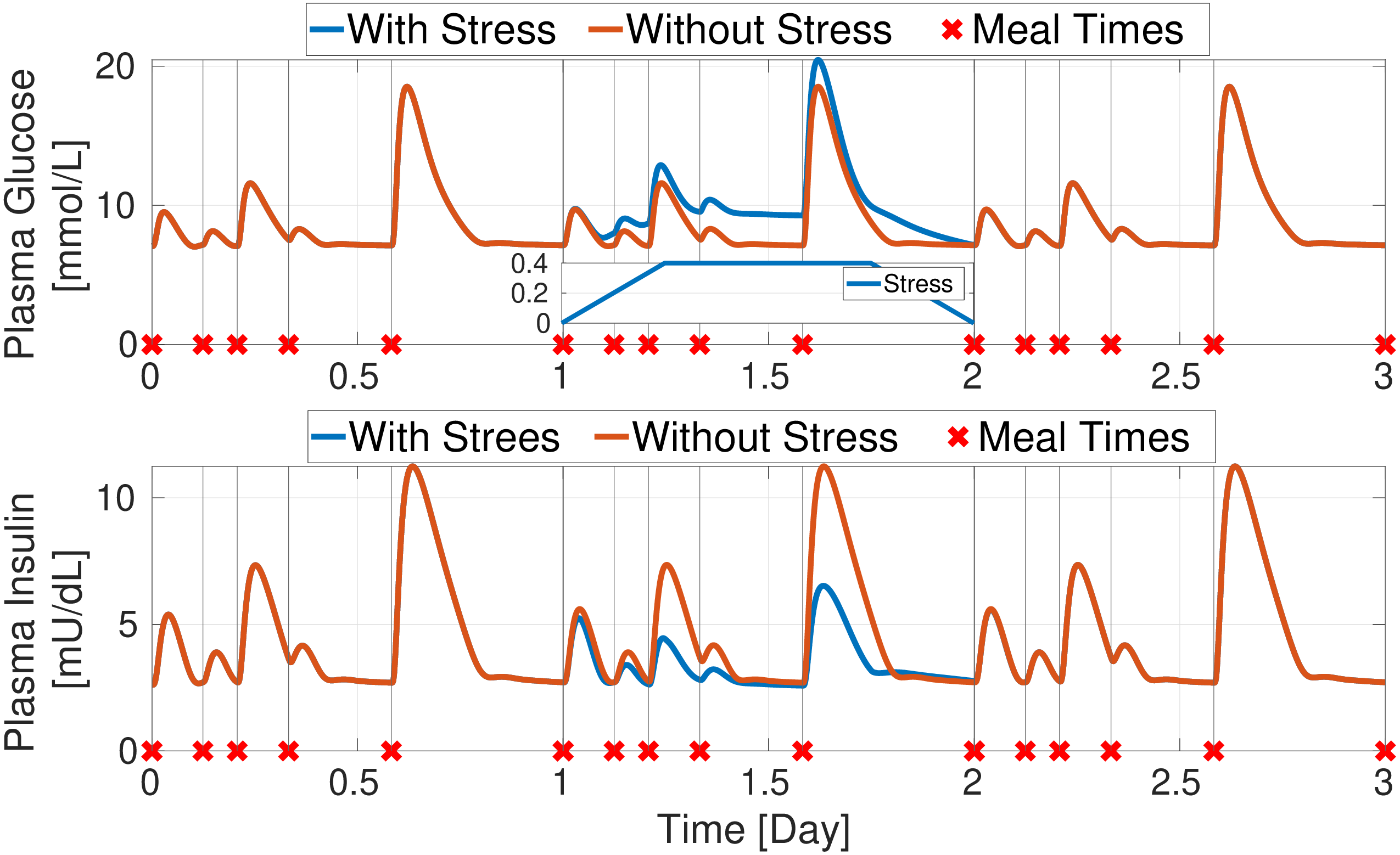}
    \caption{A simulation showing the effect of stress on glucose and insulin concentrations.}
    \label{fig:Stressplot}
\end{figure}

In this section, the effect of stress is included in the model \cite{Vahili}.
In \cite{Stress}, the effect of stress was included as a multiplicative factor $1+\alpha_{s}$, with $\alpha_{s}\in[0,1]$, to the glucose and glucagon production rates. This is based on the fact that stress causes a direct increase in the pancreatic glucagon production through an increase in catecholamines which in turns drives an increase of glucose production in the liver \cite{Woods1985}. In addition, stress was also included as a multiplicative factor $1-\alpha_{s}$ to the pancreatic insulin secretion rate based on \cite{Woods1985}. Similarly in this work, the effect of stress is included in the model as following: 
\begin{itemize}
    \item An increase in the plasma glucagon release rate $r_{P\Gamma R}$ in the glucagon compartment \eqref{eq:Gamma} as $\left(1+\alpha_{s}\right)r_{P \Gamma R}$.
    \item An increase in the hepatic glucose production rate $r_{HGP}$ in the glucose liver compartment \eqref{eq:GL} as $\left(1+\alpha_{s}\right)r_{HGP}$.
    \item A decrease in the pancreatic insulin release rate $r_{PIR}$ in the insulin liver compartment \eqref{eq:IL} as
    $\left(1-\alpha_{s}\right)r_{PIR}$
\end{itemize}
Figure \ref{fig:Stressplot} shows the effect of stress in a simulation for the same patient discussed in section \ref{sec:Glucose_Absorption_Model} when the patient is stressed on the second day with $\alpha_{s}$ ramping up from 0 to 0.4 in 6 hours, staying at 0.4 for 12 hours, and then ramping down to 0 for the rest of the day. Stress manages to increase glucose concentration together with a decrease in insulin concentration.
\section{Simulation Results}
\label{sec:SimR}
\begin{figure}[h!]
    \centering
    \includegraphics[width=0.47\textwidth]{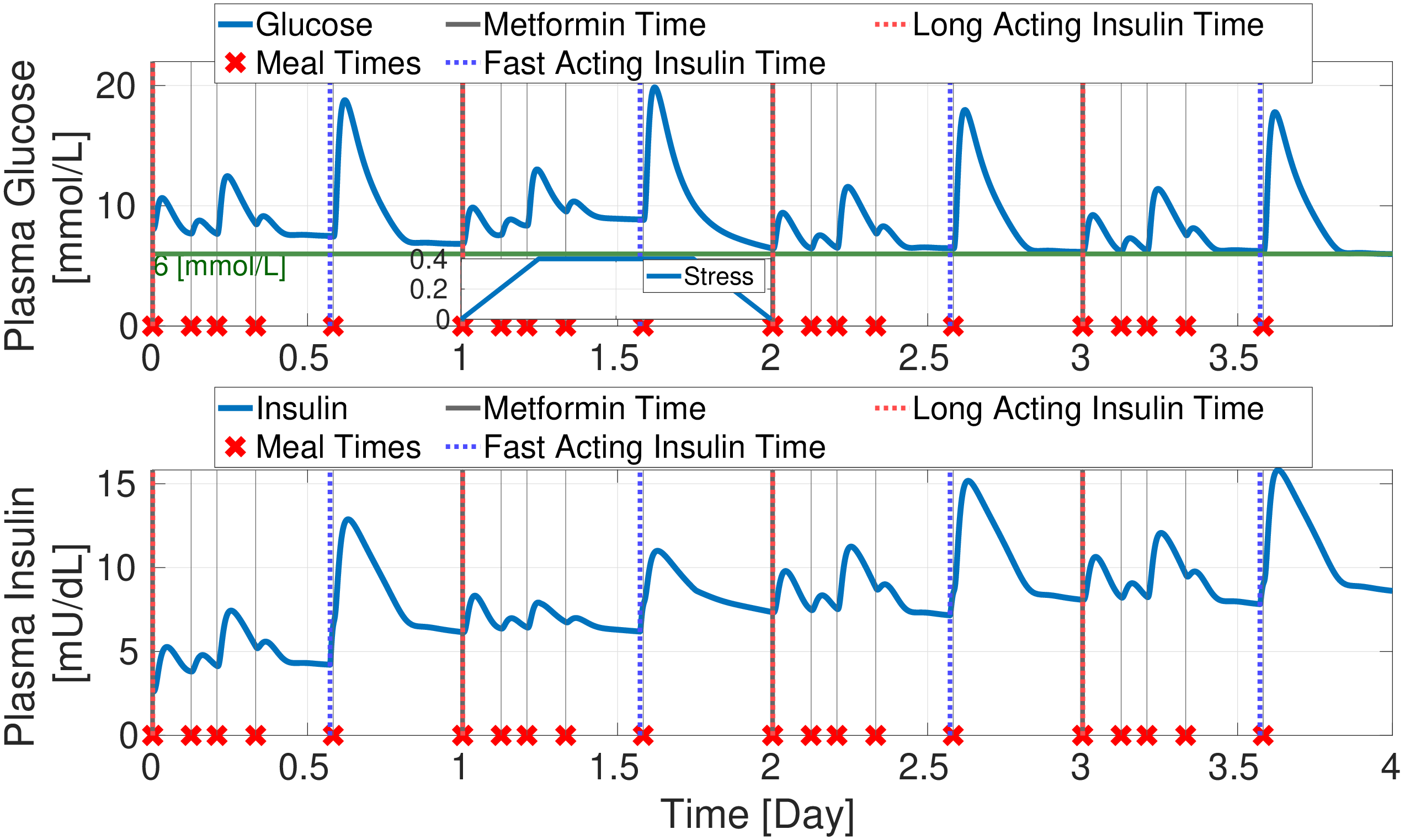}
    \caption{Simulation results for a the first case patient.}
    \label{fig:fcp}
\end{figure}

In this section, two simulation cases are carried out and discussed briefly to show how the model incorporates the effect of lifestyle changes on type 2 diabetic patients. The first case is for a patient having the same basal values and meal plan as the one discussed in section \ref{sec:Glucose_Absorption_Model}. Additionally, the patient experiences stress on the second day in the same manner as for the patient discussed in section \ref{sec:Sress}. Moreover, the patient takes a 1000 $\units{mg}$ oral dose of metformin together with long acting insulin dose of $50~\units{U}$ everyday an hour before breakfast. Furthermore, the patient take a fast acting insulin dose of $30~\units{U}$ 15 minutes before dinner. Figure \ref{fig:fcp} shows the results of the simulation for this case. The patient manages to reach a fasting plasma glucose concentration of $6~\units{mmol\,L^{-1}}$ towards the end of the third day.

\begin{figure}[h!]
    \centering
    \includegraphics[width=0.47\textwidth]{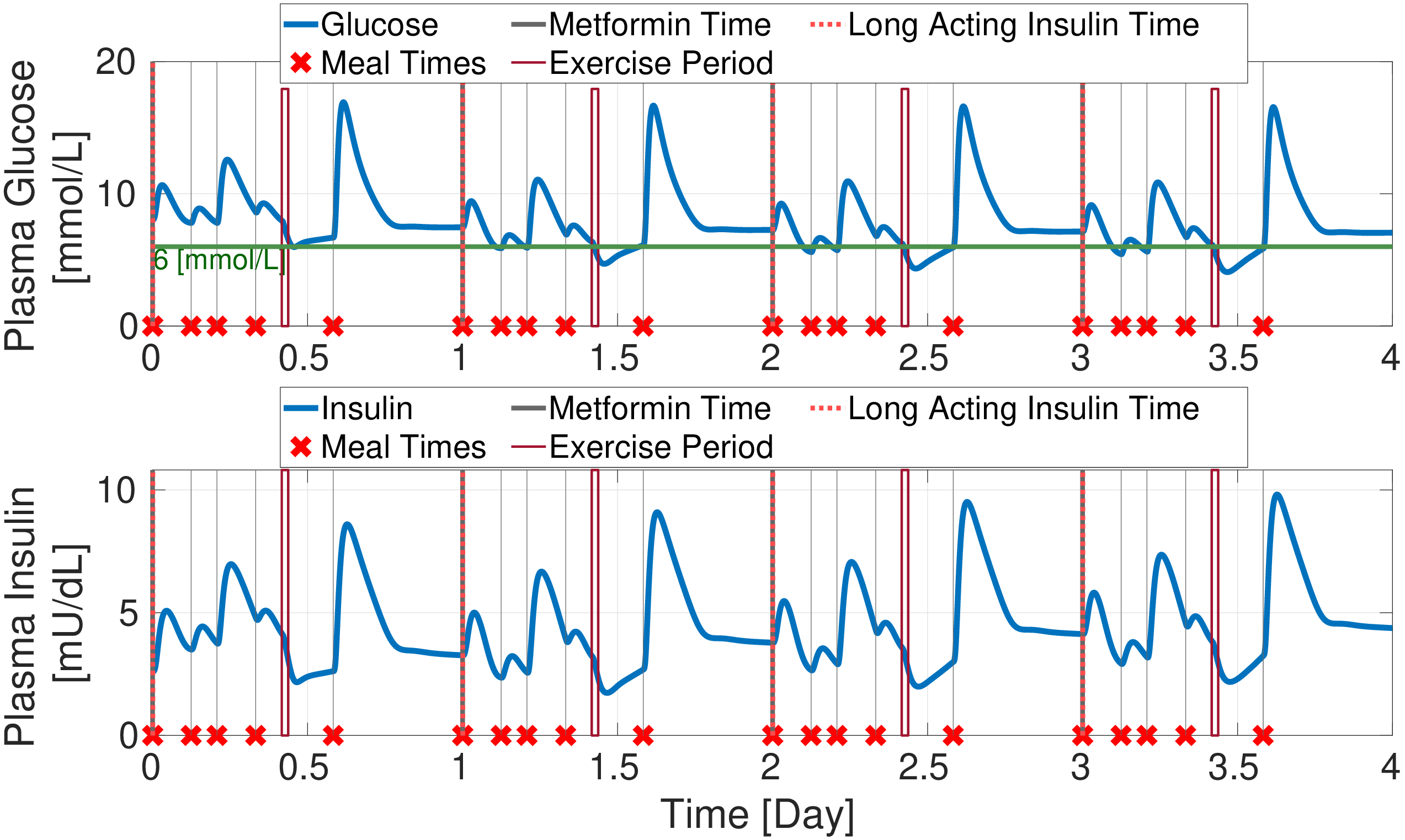}
    \caption{Simulation results for a the second case patient.}
    \label{fig:scp}
\end{figure}

For the second simulation case, the patient has the same basal values and meal plan as the first case patient. However, the patient does not experience stress and perform moderate exercise everyday raising the heart rate from 80 $\units{bpm}$ to 120 $\units{bpm}$ for a period of 30 minutes four hours before dinner. Additionally, the patient only takes a long acting insulin dose of 30 $\units{U}$ and an oral metformin dose of 500 $\units{mg}$ an hour before breakfast. Figure \ref{fig:scp} shows the second case simulation results.
While the patient does not achieve a fasting plasma glucose concentration of 6 $\units{mmol\,L^{-1}}$ by the end of the third day, the patient in the second case has lower plasma glucose concentrations during the day when compared to the first patient. 
\section{Conclusion and Future Work}
The proposed model is shown to be able to be used with multiple glucose meal sizes, different metformin doses, physical activity, insuling injections, and stress. The model, however, need to be confirmed with real patients data. Moreover, real patient data can be used to estimate joint probability distribution to model a population of T2D patients. 
\bibliographystyle{plain}        
\bibliography{autosam}           



\appendix
\section{Full Model Equations}    
\label{sec:FullModelEq}
The compartments include metabolic production rates $r_{CXP}$ and metabolic uptake rates $r_{CXU}$ for substance $X$ in compartment $C$ generally defined as following:
\begin{equation}
\label{eq:rdefine}
r_{CXP,U}=M^IM^GM^\Gamma r_{CXP,U}^b
\end{equation}
Where $M^I$,$M^G$, and $M^\Gamma$ are multiplicative quantities for the effect of insulin $I$, glucose $G$, and glucagon $\Gamma$ respectively, and $r_{CXP,U}^b$ is the basal metabolic rate of substance $X$ in compartment $C$. The general form for the multiplicative quantities representing the effect of a substance $Y$ in compartment $C$ with concentration $Y_{C}$ is given as: 
\begin{equation}
M^{Y}=\frac{a+b \tanh \left[c\left(Y_{C} / Y_{C}^{b}-d\right)\right]}{a+b \tanh [c(1-d)]}
\end{equation}
Where $Y_{C}^{b}$ is the basal concentration of substance $Y$ in compartment $C$, and $a,b, c,$ and $d$ are model parameters.
\subsection{Glucose Sub-Model}
Applying mass balance equations over the compartments for glucose, the following equations are obtained:
\begin{subequations}
\label{eq:Glucose_subsystem}
\begin{align}
\label{eq:GBC}
&V_{B C}^{G} \frac{d G_{B C}}{d t}=Q_{B}^{G}\left(G_{H}-G_{B C}\right)-\frac{V_{B F}^{G}}{T_{B}^{G}}\left(G_{B C}-G_{B F}\right) \\
\label{eq:GBF}
&V_{B F}^{G} \frac{d G_{B F}}{d t}=\frac{V_{B F}^{G}}{T_{B}^{G}}\left(G_{B C}-G_{B F}\right)-r_{B G U} \\
\label{eq:GH}
\begin{split}&V_{H}^{G} \frac{d G_{H}}{d t}=Q_{B}^{G} G_{B C}+Q_{L}^{G} G_{L}+Q_{K}^{G} G_{K}\\
&\qquad+Q_{P}^{G} G_{P C}+Q_{H}^{G} G_{H}-r_{R B C U}\end{split} \\
\label{eq:GG}
&V_{G}^{G} \frac{d G_{G}}{d t}=Q_{G}^{G}\left(G_{H}-G_{G}\right)-r^{m}_{G G U}+Ra \\
\label{eq:GL}
\begin{split}&V_{L}^{G} \frac{d G_{L}}{d t}=Q_{A}^{G} G_{H}+Q_{G}^{G} G_{G}-Q_{L}^{G} G_{L}\\
&\qquad+\left((1+\alpha_{s})\left(1-\alpha_{e}E_{2}\right)r^{m}_{H G P}-\left(1+\alpha_{e}E_{2}\right)r_{H G U}\right)\end{split} \\
\label{eq:GK}
&V_{K}^{G} \frac{d G_{K}}{d t}=Q_{K}^{G}\left(G_{H}-G_{K}\right)-r_{K G E} \\
\label{eq:GPC}
&V_{P C}^{G} \frac{d G_{P C}}{d t}=Q_{P}^{G}\left(G_{H}-G_{P C}\right)-\frac{V_{P F}^{G}}{T_{P}^{G}}\left(G_{P C}-G_{P F}\right) \\
\label{eq:GPF}
\begin{split}&V_{P F}^{G} \frac{d G_{P F}}{d t}=\frac{V_{P F}^{G}}{T_{P}^{G}}\left(G_{P C}\right.\\
&\qquad\left.-\left(1+\beta_{e}E_{1}\right)G_{P F}\right)-\left(1+\alpha_{e}E_{2}\right)r^{m}_{P G U}\end{split}
\end{align}
\end{subequations}
Where $G_{i}~\si{[mg\,dL^{-1}]}$ is glucose concentration for each compartment $i$, $Q^{G}_{i}~\si{[dL\,min^{-1}]}$ is the vascular blood flow for the glucose compartment $i$, $V^G_{i}~\si{[dL]}$  is the volume of compartment $i$, $T^G_{i}~~\si{[min]}$ is the transcapillary diffusion time for compartment $i$, and $r_{xP}, r_{xU}$ are metabolic glucose production and uptake rates respectively. The following are the meanings of each subscript in the model:
\begin{itemize}
    \item B: Brain
    \item BC: Brain capillary space
    \item BF: Brain interstitial fluid
    \item H: Hear
    \item G: Guts
    \item L: Liver
    \item K: Kidney
    \item P: Periphery
    \item PC: Periphery capillary space
    \item PF: Periphery interstitial fluid
    \item BGU: Brain glucose uptake
    \item RBCU: Red blood cell glucose uptake
    \item GGU: Gut glucose uptake
    \item HGP: Hepatic glucose production
    \item HGU: Hepatic glucose uptake
    \item KGE: Kidney glucose excretion
    \item PGU: Peripheral glucose uptake
\end{itemize}
The metabolic rates for the glucose subsystem are given as:
\begin{subequations}
\label{eq:Glucose_rates}
\begin{align}
&r_{PGU}=M^{I}_{PGU}M^{G}_{PGU}r^{b}_{PGU}, ~ r^{b}_{PGU}=35\\
&r_{HGP}=M^{I}_{HGP}M^{G}_{HGP}M^{\Gamma}_{HGP}r^{b}_{HGP},~r^{b}_{HGP}=35\\
&r_{HGU}=M^{I}_{HGU}M^{G}_{HGU}r^{b}_{HGU}, ~ r^{b}_{HGU}=20\\
&r_{KGE}=\begin{cases}
71+71\tanh\left[0.11\left(G_{K}-460\right)\right] & G_{K} < 460\\
-330+0.872G_{K} & G_{K}\geq460\end{cases}\\
&r_{BGU}=70,~r_{RBCU}=10,~r_{GGU}=20
\end{align}
\end{subequations}
Where:
\begin{subequations}
\label{eq:Glucose_ratesMF}
\begin{align}
&M_{PGU}^{I}=\frac{7.03+6.52 \tanh \left[c_{1}\left(I_{PF} / I_{PF}^{B}-d_{1}\right)\right]}{7.03+6.52 \tanh [c_{1}(1-d_{1})]}\\
&M^{G}_{PGU}=G_{PF}/G^{b}_{PF}\\
&\frac{d}{d t} M_{HGP}^{I}=0.04\left(M_{HGP}^{I \infty}-M_{HGP}^{I}\right)\\
&M_{HGP}^{I \infty}=\frac{1.21-1.14 \tanh \left[c_{2}\left(I_{L} / I_{L}^{B}-d_{2}\right)\right]}{1.21-1.14 \tanh [c_{2}(1-d_{2})]}\\
&M_{HGP}^{G}=\frac{1.42-1.41 \tanh \left[c_{3}\left(G_{L} / G_{L}^{B}-d_{3}\right)\right]}{1.42-1.41 \tanh [c_{3}(1-d_{3})]}\\
&M_{HGP}^{\Gamma}=2.7 \tanh \left[0.39 \Gamma / \Gamma^{B}\right]-f\\
&\frac{d}{\mathrm{d} t} f=0.0154\left[\left(\frac{2.7 \tanh \left[0.39 \Gamma /\Gamma^{B}\right]-1}{2}\right)-f\right]\\
&\frac{d}{d t} M_{HGU}^{I}=0.04\left(M_{HGU}^{I \infty}-M_{HGU}^{I}\right)\\
&M_{HGU}^{I \infty}=\frac{2.0 \tanh \left[c_{4}\left(I_{L} / I_{L}^{B}-d_{4}\right)\right]}{2.0 \tanh [c_{4}(1-d_{4})]}\\
&M_{HGU}^{G}=\frac{5.66+5.66 \tanh \left[c_{5}\left(G_{L} / G_{L}^{B}-d_{5}\right)\right]}{5.66+5.66 \tanh [c_{5}(1-d_{5})]}
\end{align}
\end{subequations}
Note that some of these rates have a constant numerical value. In addition, parameters $a$ and $b$ for the multiplicative quantities are substituted with numerical values. These numerical values are the ones estimated for a healthy 70 \si{kg} male. Parameters $c$ and $d$ were left for the estimation in case of a diabetic patient as in \cite{Vahidi2016}.
The following rates are modified with the effect of metformin as following:
\begin{subequations}
\label{eq:Modefied_rates_Metaformin}
\begin{align}
    \label{eq:rmGGU}
    r^{m}_{GGU}&=\left(1+E_{GW}\right)r_{GGU}\\
    \label{eq:rmHGP}
    r^{m}_{HGP}&=\left(1-E_{L}\right)r_{HGP}\\
    \label{eq:rm}
    r^{m}_{PGU}&=\left(1+E_{P}\right)r_{PGU}
\end{align}
\end{subequations}
Where $E_{GW}$, $E_{L}$, and $E_{P}$ are positives coefficients which depend on the amount of metformin in the gastrointestina wall (GI) $M_{GW}~\units{\mu g}$, liver $M_{L}~\units{\mu g}$, and peripherals $M_{p}~\units{\mu g}$ respectively. These coefficients increase (or decrease) the glucose uptake (or production) as seen in \eqref{eq:Modefied_rates_Metaformin}.
The equations for these coefficients are given as following:
\begin{subequations}
\label{eq:Metaformin_coef}
\begin{align}
&E_{G W}=\frac{\nu_{G W, \max } \times\left(M_{GW}\right)^{n_{G W}}}{\left(\varphi_{G W, 50}\right)^{n_{G W}}+\left(M_{GW}\right)^{n_{G W}}}\\
&E_{L}=\frac{\nu_{L, \max } \times\left(M_{L}\right)^{n_{L}}}{\left(\varphi_{L, 50}\right)^{n_{L}}+\left(M_{L}\right)^{n_{L}}}\\
&E_{P}=\frac{\nu_{P, \max } \times\left(M_{P}\right)^{n_{P}}}{\left(\varphi_{P, 50}\right)^{n_{P}}+\left(X_{P}\right)^{n_{P}}}
\end{align}
\end{subequations}
Where $\nu_{GW, \max }$, $\nu_{L, \max }$, $\nu_{P, \max }$ are parameters to represent the maximum effect of metformin in each one of its corresponding compartments, $\varphi_{G W, 50}$, $\varphi_{G I, 50}$, $\varphi_{G I, 50}~\units{\mu g}$ are the masses of metformin within the different compartments to produce half of its maximum effect, and $n_{G W},~n_{L}$, and $n_{P}$ are shape factors.
\subsection{Incretins Sub-Model}
The incretins hormones are metabolic hormones released after eating a meal to stimulate a decrease in blood glucose levels. For T2D patients, Glucagon-Like-Peptide-1 (GLP-1) is the most active incretin \cite{Garber2011}. GLP-1 is then modelled with the following two compartments as in \cite{Vahili}:
\begin{subequations}
\label{eq:GLP1}
\begin{align}
&\frac{d\psi}{dt}=\zeta k_{\mathrm{empt}}q_{Sl}-\frac{1}{\tau_{\psi}}\psi\\
&V^{\Psi}\frac{d\Psi}{dt}=\frac{1}{\tau_{\psi}}\psi-\left[K_{out}+(R_{maxC}-DR_{c})\mathit{Cf}_{2}\right]\Psi
\end{align}
\end{subequations}

Where $\tau_{\psi}~[\si{min^{-1}}]$ is a time constant for the release and absorption of GLP-1 to the blood stream upon consuming a meal, $V^{\Psi}~[\si{dL}]$ is the volume of the GLP-1 compartment, $DR_{c}~[\si{nmol}]$ is the amount of Dipeptidyl peptidase-4 (DPP-4) in the central compartment deactivated by the drug vildagliptin,  $K_{out}~[\si{min^{-1}}]$ is a clearance constant for GLP-1 independently of the amount of DPP-4, and  $(R_{maxC}-DR_{c})$ is the amount of available activated DPP-4 in the blood plasma with $R_{maxC}~[\si{nmol}]$ being the maximum amount of active DPP-4 in the absence of the vildagliptin. $\mathit{Cf}_{2}~[\si{min^{-1}nmol^{-1}}]$ is a proportionality factor for the elimination of GLP-1 by active DPP-4 \cite{Landersdorfer2012}. Parameters $\tau_{\psi}$ and $\zeta$ are estimated in \cite{Vahidi2016} when other incretins than GLP-1 are considered and later modified in \cite{Vahidi2016} to account for vildagliptin treatment. Parameters $K_{out}$, $R_{maxC}$, and $\mathit{Cf}_{2}$ were estimated in \cite{Landersdorfer2012} together with the parameters for the vildagliptin model described in subsection \ref{sec:Vilmodel}.
\subsection{Glucagon Sub-Model}
The glucagon subsystem consists of one compartment as it is assumed to have the same concentration over all the body:
\begin{subequations}
    \label{eq:Glucagon}
    \begin{align}
    \label{eq:Gamma}
    &V^{\Gamma}\frac{d\Gamma}{dt}=(1+\alpha_{s})r_{P \Gamma R}-9.1\Gamma\\
    &r_{P \Gamma R}=M^{G}_{P \Gamma R}M^{I}_{P \Gamma R}r^{b}_{P \Gamma R},~ r^{b}_{P \Gamma R}=9.1\\
    &M_{P \Gamma R}^{G}=1.31-0.61 \tanh \left[1.06\left(\frac{G_{H}}{G_{H}^{B}}-0.47\right)\right]\\
    &M_{P \Gamma R}^{I}=2.93-2.09 \tanh \left[4.18\left(\frac{I_{H}}{I_{H}^{B}}-0.62\right)\right]
    \end{align}
\end{subequations}
Where $r_{P \Gamma R}$ is the plasma glucagon release rate. The state $\Gamma$ represent a normalized glucagon state with respect to its basal value. This is done since it is difficult in practice to obtain glucagon measurements for each subject in order to initialize the state. Therefore for this model, the basal glucagon state is 1.

\subsection{Insulin Sub-Model}
Applying mass balance equations over the insulin compartments will yield the following: 
\begin{subequations}
\label{eq:Insulin_subsystem}
\begin{align}
\label{eq:IB}
&V_{B}^{I} \frac{d I_{B}}{d t}=Q_{B}^{I}\left(I_{H}-I_{B}\right)\\
\label{eq:IH}
\begin{split}&V_{H}^{I} \frac{d I_{H}}{d t}=Q_{B}^{I} I_{B}+Q_{L}^{I} I_{L}+Q_{K}^{I} I_{K}\\
&\qquad+Q_{P}^{I} I_{P V}-Q_{H}^{I} I_{H}\end{split}\\
\label{eq:IG}
&V_{G}^{I} \frac{d I_{G}}{d t}=Q_{G}^{I}\left(I_{H}-I_{G}\right)\\
\label{eq:IL}
\begin{split}&V_{L}^{I} \frac{d L_{L}}{d t}=Q_{A}^{I} I_{H}+Q_{G}^{I} I_{G}-Q_{L}^{I} I_{L}\\&\qquad+(1-\alpha_{s})r_{PIR}-r_{LIC}\end{split}\\
\label{eq:IK}
&V_{K}^{I} \frac{d I_{K}}{dt}=Q_{K}^{I}\left(I_{H}-I_{K}\right)-r_{KIC}\\
\label{eq:IPC}
&V_{PC}^{I} \frac{d I_{P C}}{d t}=Q_{P}^{I}\left(I_{H}-I_{PC}\right)-\frac{V_{PF}^{I}}{T_{P}^{I}}\left( I_{PC}-I_{PF}\right)\\
\label{eq:IPF}
&V_{PF}^{I} \frac{dI_{PF}}{d t}=\frac{V_{PF}^{I}}{T_{P}^{I}}\left(I_{PC}-I_{PF}\right)-r_{PIC}+r_{Inj}
\end{align}
\end{subequations}
Where $r_{LIC}$, $r_{KIC}$, and $r_{PIC}$ are the liver, kidney, and peripherals insulin clearance rates respectively and are defined as following:
\begin{subequations}
\label{eq:rLIC,rKIC}
\begin{align}
&r_{LIC}=0.4\left[Q_{A}^{I} I_{H}+Q_{G}^{I} I_{G}-Q_{L}^{I} I_{L}+r_{PIR}\right]\\
&r_{KIC}=0.3Q_{K}^{I}I_{K}\\
&r_{PIC}=\frac{I_{PF}}{\left[\left(\frac{1-0.15}{0.15 Q_{P}^{I}}\right)-\frac{20}{V_{PF}^{I}}\right]}
\end{align}
\end{subequations}
the pancreas insulin release is calculated by the following:
\begin{equation}
    \label{eq:rPIR}
    r_{PIR}=\frac{S}{S^{b}}r^b_{PIR}
\end{equation}
Where $S~[\si{U\,min^{-1}}]$ is the pancreas secreted insuln rate, and $S^{b}$, $r^{b}_{PIR}$ are the basal values. The model for $S$ and $S^{b}$ is described in subsection \ref{sec:PancModel}.

\subsection{Pancreas Sub-Model}
\label{sec:PancModel}
The model consists of two main compartments: a large insulin storage compartment $m_{s}~\units{\mu g}$ and a small labile insulin compartment $m_{l}~\units{\mu g}$. The flow of insulin from the storage compartment to the labile insulin compartment is dependent on a dimensionless factor $P$ with a proportionality constant $\gamma~\units{\mu g\,min^{-1}}$. The factor $P$ depends on a dimensionless glucose-enhanced excitation factor represented by $X$ and GLP-1 through a linear compartment with constant first order rate $\alpha~\units{min^{-1}}$. Upon a glucose stimulus, the glucose-enhanced excitation factor $X$ will increase instantaneously depending on the glucose increase in the plasma $G_{H}$. In addition, a dimensionless inhibitor $R$ for $X$ will increase in response to $X$ through a linear compartment with a first order constant rate $\beta~\units{min^{-1}}$. During that increase, the secreted insulin  $S$ will depend directly on both $X$ and its inhibitor $R$ together with GLP-1. Afterwards when $R$ reaches $X$ or $X$ starts decreasing after $R$ reaching it, the insulin secretion rate will only depend on $X$ and GLP-1. 
The following are the equations of the model:
\begin{subequations}
\label{eq:Pancreas_model}
\begin{align}
    \label{eq:ms}
    &\frac{dm_{s}}{dt}=K_{l}m_{l}-K_{s}m_{s}-\gamma P\\
    \label{eq:ml}
    &\frac{dm_{l}}{dt}=K_{s}m_{s}-K_{l}m_{l}+\gamma P-S\\
    \label{eq:P}
    &\frac{dP}{dt}=\alpha(P_{\infty}-P)\\
    \label{eq:R}
    &\frac{dR}{dt}=\beta(X-R)\\
    \label{eq:S}
    &S=\begin{cases}
    \left[N_{1}P_{\infty}+N_{2}\left(X-R\right)+\zeta_{2}\Psi\right]m_{l} & X>R\\
    \left(N_{1}P_{\infty}+\zeta_{2}\Psi \right)m_{l} & X\leq R
    \end{cases}\\
    \label{eq:Pinfty}
    &P_{\infty}=X^{1.11}+\zeta_{1}\Psi\\
    \label{eq:X}
    X&=\frac{G_{H}^{3.27}}{132^{3.27}+5.93 G_{H}^{3.02}}
\end{align}
\end{subequations}
Where $K_{l}~[\si{min^{-1}}]$ and $K_{s}~[\si{min^{-1}}]$ are the rates for the flow between the labile and storage insulin compartments independently of $P$, $N_{1}~[\si{min^{-1}}]$ and $N_{2}~[\si{min^{-1}}]$ are constant parameters that represent the effect of $P$ and $\left(X-R\right)$ on the insulin secretion rate respectively, and $\zeta_{1}~\units{L\,pmol^{-1}}$, $\zeta_{2}~\units{L\,(pmol\,min)^{-1}}$ are constant parameters to represent the effect of GLP-1 on $P_{\infty}$ and the insulin secretion rate. For initializing the model and calculating the basal values, the storage compartment is assumed to be large enough for it to be constant. Therefore, writing the mass balance for the storage compartment at zero glucose concentration will yield the following:
\begin{equation}
    \label{eq:SSms}
    K_{s}m_{s}=K_{l}m_{l_{0}}
\end{equation}
Where $m_{l_{0}}$ is the labile insulin concentration at zero glucose concentration. This parameter in \cite{sorensen1985physiologic} is provided with a value of $6.33~\si{[U]}$ for a healthy $70~\si{kg}$ male.
\subsection{Vildagliptin}
\label{sec:Vilmodel}
The vildagliptin model is based on \cite{Landersdorfer2012}. The absorption of orally ingested vildagliptin is modelled by two compartments as following:
\begin{subequations}
\label{eq:VildagliptinAG}
\begin{align}
\label{eq:AG1}
&\frac{dA_{G1}}{dt}=-k_{a1}A_{G1}+\sum_{i=1}^{N_{v}(t)}\delta\left(t-t_{i}\right)f_{v}u_{v_{i}}\\
&\frac{d A_{G 2}}{d t}=k_{a 1} \times A_{G 1}-k_{a 2} \times A_{G 2}
\end{align}
\end{subequations}
Where $A_{G1},~A_{G2}~\units{nmol}$ are the amount of vildagliptin in the gut and absorption compartments respectively, $N_{v}(t)$ is the number of oral vildagliptin doses until time $t$, $u_{v_{i}}~\units{nmol}$ is the amount of consumed vildagliptin, $f_{a}$ is the bioavailability of vlidagliptin, and $k_{a1},~k_{a2}~\units{min^{-1}}$ are rate absorption parameters. After that, the model contains a central and a peripheral compartment for the vildagliptin and the vildagliptin-DPP-4 complex (deactivated DPP-4):
\begin{subequations}
\label{eq:Vildagliptin_modelCP}
\begin{align}
    \label{eq:AC}
\begin{split}
&\frac{d A_{c}}{d t}=k_{a 2}A_{G 2}-\frac{\mathit{CL}+\mathit{CL}_{\mathit{ic}}}{V_{C}}A_{C}+\frac{\mathit{CL}_{\mathit{ic}}}{V_{p}}A_{p}\\&\qquad-
\frac{\left(R_{\max C}-D R_{C}\right)k_{v2}\frac{A_{c}}{V_{C}}}{K_{vd}+\frac{A_{c}}{V_{c}}}+k_{o f f} D R_{C}
\end{split}\\
\label{eq:AP}
\begin{split}&\frac{d A_{p}}{d t}=\mathit{CL}_{\mathit{ic}} \left(\frac{A_{c}}{V_{c}}-\frac{A_{p}}{V_{p}}\right)\\
&\qquad-\frac{\left(R_{\max P}-D R_{p}\right) k_{v2} \frac{A_{p}}{V_{p}}}{K_{vd}+\frac{A_{p}}{V_{p}}}+k_{o f f}D R_{p}\end{split}\\
\begin{split}&\frac{d D R_{C}}{d t}=\frac{\left(R_{\max C}-D R_{C}\right) k_{v2} \frac{A_{C}}{V_{C}}}{K_{vd}+\frac{A_{C}}{V_{C}}}\\
&\qquad-\left(k_{off}+k_{deg}\right) D R_{C}\end{split}\\
\begin{split}&\frac{d D R_{P}}{d t}=\frac{\left(R_{\max P}-D R_{P}\right) k_{v2} \frac{A_{P}}{V_{P}}}{K_{vd}+\frac{A_{P}}{V_{P}}}\\
&\qquad-\left(k_{off}+k_{deg}\right) D R_{P}\end{split}
\end{align}
\end{subequations}
Where $A_{C},~A_{P}~\units{nmol}$ are the amounts of vildagliptin in the central and peripheral compartments respectively, $\mathit{CL}~\units{L\,min^{-1}}$ is a non-saturable clearance, $\mathit{CL}_{\mathit{ic}}~\units{L\,{min}^{-1}}$ is the inter-compartmental clearance, $V_{c},~V_{p}~\units{L}$ are the volumes of the central and peripheral compartments respectively,  $k_{v2}~[\si{min^{-1}}]$ is a parameter added for the slow tight binding of vildagliptin to DPP-4, $K_{vd}~\units{nmol\,L^{-1}}$ is the equilibrium dissociation constant, $k_{off}~\units{min^{-1}}$ is a rate constant for the dissociation of intact vildaglptin from DPP-4, $R_{maxP}~\units{nmol}$ is the maximum possible amount of DPP-4 in the peripheral compartment, $k_{deg}~\units{min^{-1}}$ is a rate constant for the hydrolysis of vildagliptin by DPP-4, and $DR_{P}~\units{nmol}$ is the amount of deactivated DPP-4 in the peripheral compartments.
\section{Parameters Mean Values}
\label{sec:param}
\begin{table*}[t]
    \centering
    \begin{tabular}{c|c||c|c||c|c}
    \hline
    Parameter & Value & Parameter & Value & Parameter & Value\\
    \hline
         $V^{G}_{BC}~\si{[dL]}$ & 3.5 & $V^{G}_{BF}~\si{[dL]}$ & 4.5 & $V^{G}_{H}~\si{[dL]}$ & 13.8 \\
         \hline
         $V^{G}_{L}~\si{[dL]}$ & 25.1 &
         $V^{G}_{G}~\si{[dL]}$ & 11.2 & $V^{G}_{K}~\si{[dL]}$ & 6.6 \\
         \hline
         $V^{G}_{PC}~\si{[dL]}$ & 10.4 &
         $V^{G}_{PF}~\si{[dL]}$ & 67.4 &
         $V^{I}_{B}~\si{[L]}$ & 0.26 \\
         \hline
         $V^{I}_{H}~\si{[L]}$ & 0.99 & $V^{I}_{G}~\si{[L]}$ & 0.94 &  $V^{I}_{L}~\si{[L]}$ & 1.14 \\
         \hline
         $V^{I}_{K}~\si{[L]}$ & 0.51 & $V^{I}_{PC}~\si{[L]}$ & 0.74 & $V^{I}_{PF}~\si{[L]}$ & 6.74 \\
         \hline
         $V^{\Gamma}~\si{[mL]}$ & 6.74 &
         $Q^{G}_{B}~\si{[dL\,min^{-1}]}$ & 5.9 & $Q^{G}_{H}~\si{[dL\,min^{-1}]}$ & 43.7 \\
         \hline 
         $Q^{G}_{A}~\si{[dL\,min^{-1}]}$ & 2.5 & $Q^{G}_{L}~\si{[dL\,min^{-1}]}$ & 12.6 &
         $Q^{G}_{G}~\si{[dL\,min^{-1}]}$ & 10.1 \\
         \hline
         $Q^{G}_{K}~\si{[dL\,min^{-1}]}$ & 10.1 & $Q^{G}_{P}~\si{[dL\,min^{-1}]}$ & 15.1 & $Q^{I}_{B}~\si{[dL\,min^{-1}]}$ & 0.45 \\
         \hline
         $Q^{I}_{H}~\si{[L\,min^{-1}]}$ & 3.12 & $Q^{I}_{A}~\si{[L\,min^{-1}]}$ & 0.18 & 
         $Q^{I}_{K}~\si{[L\,min^{-1}]}$ & 0.72 \\
         \hline
         $Q^{I}_{P}~\si{[L\,min^{-1}]}$ & 1.05 & $Q^{I}_{G}~\si{[L\,min^{-1}]}$ & 0.72 & $Q^{I}_{L}~\si{[L\,min^{-1}]}$ & 0.9\\
         \hline
         
         $T^{G}_{B}~\si{[min]}$ & 2.1 & $T^{G}_{P}~\si{[min]}$ & 5.0 & 
         $T^{I}_{P}~\si{[min]}$ & 20.0 \\
         \hline
         $f_{q}~\units{\cdot}$ & 0.9 &
         $k_{\phi1}~\units{\cdot}$ & 0.68 & $k_{\phi2}~\units{\cdot}$ & 0.00236 \\
         \hline
         $k_{12_{q}}~\si{[min^{-1}]}$ & 0.08 & $k_{\mathrm{min}}~\si{[min^{-1}]}$ & 0.005 & $k_{\mathrm{max}}~\si{[min^{-1}]}$ & 0.05 \\
         \hline
         $k_{\mathrm{abs}}~\si{[min^{-1}]}$ & 0.08 &
         $c_{1}~\units{\cdot}$ & 0.067 & $c_{2}~\units{\cdot}$ & 1.59\\
         \hline
         $c_{3}~\units{\cdot}$ & 0.62 & $c_{4}~\units{\cdot}$ & 1.72 &
         $c_{5}~\units{\cdot}$ & 2.03 \\
         \hline
         $d_{1}~\units{\cdot}$ & 1.126 & $d_{2}~\units{\cdot}$ & 0.683 & $d_{3}~\units{\cdot}$ & 0.14 \\
         \hline
         $d_{4}~\units{\cdot}$ & 0.023 & $d_{5}~\units{\cdot}$ & 1.59 & $m_{l_{0}}~\si{[U]}$ & 6.33 \\
         \hline
         $\zeta_{1}~\si{[L\,pmol^{-1}]}$ & 0.0026 &
         $\zeta_{2}~\si{[L\,(pmol\,min)^{-1}]}$ & $0.99\text{e}^{-4}$ & $K_{l}~[\si{min^{-1}}]$ & 0.3621 \\
         \hline
         $K_{s}~[\si{min^{-1}}]$ & 0.0572 & $\gamma~\units{\mu g\,min^{-1}}$ & 2.366 & $\alpha~\units{min^{-1}}$ & 0.615 \\
         \hline
         $\beta~\units{min^{-1}}$ & 0.931 & $N_{1}~\units{min^{-1}}$ & 0.0499 & $N_{2}~\units{min^{-1}}$ & 0.00015 \\
         \hline
         $V^{\Psi}~\units{dL}$ & 11.31 & $K_{out}~\units{min^{-1}}$ & 68.3041 & $\mathit{Cf}_{2}~\units{min^{-1}nmol^{-1}}$ & 21.1512 \\
         \hline
         $\tau_{\psi}~\units{min^{-1}}$ & 35.1 & $R_{maxC}~\units{nmol}$ & 5.0 & $\zeta~\units{\cdot}$ & 8.248 \\
         \hline
         $f_{v}~\units{\cdot}$ & 0.772 & $k_{a1}~\units{min^{-1}}$ & 0.021 &
         $k_{a2}~\units{min^{-1}}$ & 0.0175 \\
         \hline
         $\mathit{CL}~\units{L\,min^{-1}}$ & 0.6067 & $\mathit{CL}_{\mathit{ic}}~\units{L\,min^{-1}}$ & 0.6683 & $V_{p}~\units{L}$ & 97.3 \\
         \hline
         $k_{off}~\units{min^{-1}}$ & 0.0102 & $R_{maxP}~\units{nmol}$ & 13 & $k_{deg}~\units{min^{-1}}$ & 0.0018 \\
         \hline
         $V_{c}~\units{L}$ & 22.2 &
         $K_{vd}~\units{nmol\,L^{-1}}$ & 71.9 & $k_{v2}~\units{min^{-1}}$ & 0.39 \\
         \hline
         
         $k_{go}~\units{min^{-1}}$ & $1.88\text{e}{-4}$ & $k_{gg}~\units{min^{-1}}$ & $1.85\text{e}{-4}$ &
         $k_{pg}~\units{min^{-1}}$ & 4.13 \\
         \hline
         
         $k_{gl}~\units{min^{-1}}$ & 0.46 & $k_{pl}~\units{min^{-1}}$ & 0.00101 & $k_{lp}~\units{min^{-1}}$ & 0.91 \\
         \hline
         
         $k_{po}~\units{min^{-1}}$ & 0.51 & $\nu_{G W, \max }~\units{\cdot}$ & 0.9720 & $\nu_{L, \max }~\units{\cdot}$ & 0.7560 \\
         \hline
         
         $\nu_{P, \max }~\units{\cdot}$ & 0.2960 &
         $n_{GW}~\units{\cdot}$ & 2.0 & $n_{L}~\units{\cdot}$ & 2.0 \\
         \hline
         
         $n_{P}~\units{\cdot}$ & 5.0 & $\phi_{GW, 50 }~\units{\cdot}$ & 431.0 & $\phi_{L, 50 }~\units{\cdot}$ & 521.0 \\
         \hline
         
         $\phi_{P, 50 }~\units{\cdot}$ & 1024.0 & $\rho_{\alpha}~\units{min^{-1}}$ & 54 & $\rho_{\beta}~\units{min^{-1}}$ & 54 \\
         \hline
         
         $\alpha_{M}~\units{min^{-1}}$ & 0.06 & $\beta_{M}~\units{min^{-1}}$ & 0.1 & $p_{la}~\units{min^{-1}}$ & 0.5 \\
         \hline
         
         $r_{la}~\units{\cdot}$ & 0.2143 &
         $q_{la}~\units{dL^2\,mU^{-2}}$ & $3.04\text{e}^{-10}$ & $b_{la}~\units{min^{-1}}$ & 0.025 \\
         \hline
         
         $C_{\mathrm{max}}~\units{\cdot}$ & 15.0 & $k_{la}~\units{min^{-1}}$ & $2.35\text{e}{-5}$ &
         $p_{fa}~\units{min^{-1}}$ & 0.5 \\
         \hline
         
         $r_{fa}~\units{\cdot}$ & 0.2143 & $q_{fa}~\units{dL^2\,mU^{-2}}$ & $1.3\text{e}{-11}$ & $b_{fa}~\units{min^{-1}}$ & 0.0068 \\
         \hline
         
         $\tau_{\mathit{HR}}~\units{min}$ & 5.0 & $n_{e}~\units{\cdot}$ & 4.0 & $a_{e}~\units{\cdot}$ & 0.8 \\
         \hline
         
         $\tau_{e}~\units{min}$ & 600 &
         $\alpha_{e}~\units{\cdot}$ & 2.974 & $\beta_{e}~\units{bpm^{-1}}$ & $3.39\text{e}{-4}$\\
         \hline
         
    \end{tabular}
     \captionsetup{aboveskip=0pt,font=it}    \caption{Parameters Values}
    \label{tab:Param}
\end{table*}
 Table \ref{tab:Param} includes the values of the parameters which were used in the simulation.
\end{document}